\documentclass[twocolumn]{pasj00}
\SetRunningHead{T. Nakajima et al.}{New 100-GHz Dual-Pol. 2SB Receiver}
\Received{2008/02/01}
\Accepted{2008/03/30}
\Published{2008/**/**}

\begin{document}
\title{A New 100-GHz Band Front-End System with\\
a Waveguide-Type Dual-Polarization Sideband-Separating\\
SIS Receiver for the NRO 45-m Radio Telescope}
\author{Taku \textsc{Nakajima}\altaffilmark{1}, Masayuki \textsc{Kawamura}\altaffilmark{1}, Kimihiro \textsc{Kimura}\altaffilmark{1}, Yoshinori \textsc{Yonekura}\altaffilmark{1},\\
Hideo \textsc{Ogawa}\altaffilmark{1}, Takeshi \textsc{Sakai}\altaffilmark{2}, Nario \textsc{Kuno}\altaffilmark{2}, Masato \textsc{Tsuboi}\altaffilmark{3},\\
Shin'ichiro \textsc{Asayama}\altaffilmark{4}, Takashi \textsc{Noguchi}\altaffilmark{4}, and Ryohei \textsc{Kawabe}\altaffilmark{2}}
\altaffiltext{1}{Department of Physical Science, Graduate School of Science, Osaka Prefecture University,\\
1-1 Gakuen-cho, Naka-ku, Sakai, Osaka 599-8531}
\altaffiltext{2}{Nobeyama Radio Observatory, National Astronomical Observatory of Japan,\\
462-2 Nobeyama, Minamimaki, Minamisaku, Nagano 384-0305}
\altaffiltext{3}{Institute of Space and Astronautical Science, Japan Aerospace Exploration Agency,\\
3-1-1 Yoshinodai, Sagamihara, Kanagawa 229-8510}
\altaffiltext{4}{Advanced Technology Center, National Astronomical Observatory of Japan,\\
2-21-1 Osawa, Mitaka, Tokyo 181-8588}
\email{s\_tac@p.s.osakafu-u.ac.jp}
\maketitle

\begin{abstract}
We developed a waveguide-type dual-polarization sideband-separating SIS receiver system of the 100-GHz band for the 45-m radio telescope at the Nobeyama Radio Observatory, Japan. This receiver is composed of an ortho-mode transducer and two sideband-separating SIS mixers, which are both based on the waveguide technique. The receiver has four intermediate frequency bands of 4.0--8.0 GHz. Over the radio frequency range of 80--120 GHz, the single-sideband receiver noise temperatures are 50--100 K and the image rejection ratios are greater than 10 dB. We developed new matching optics for the telescope beam as well as new IF chains for the four IF signals. The new receiver system was installed in the telescope, and we successfully observed the $^{12}$CO, $^{13}$CO and C$^{18}$O emission lines simultaneously toward the Sagittarius B2 region to confirm the performance of the receiver system. The SSB noise temperature of the system, including the atmosphere, became approximately half of that of the previous receiver system. The Image Rejection Ratios (IRRs) of the two 2SB mixers were calculated from the $^{12}$CO and HCO$^{+}$ spectra from the W51 giant molecular cloud, resulting in $>$ 20 dB for one polarization and $>$ 12 dB for the other polarization.
\end{abstract}

\begin{figure}
  \begin{center}
  \FigureFile(80mm,60mm){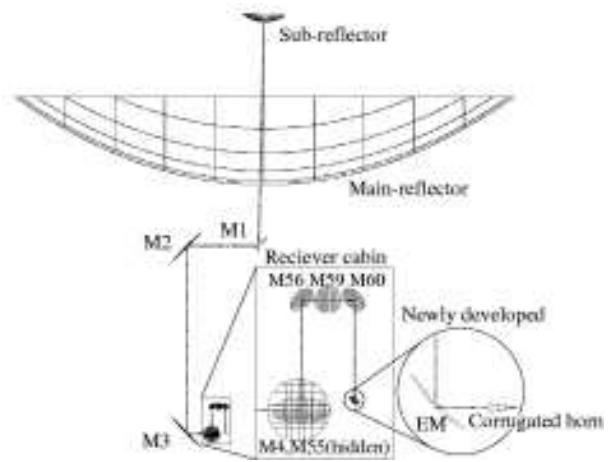}
  \end{center}
  \caption{Optics of the 45-m telescope. The developed receiver system is placed in the receiver cabin. EM denotes an ellipsoidal mirror.}\label{fig1}
\end{figure}

\begin{table}
\begin{center}
\caption{Physical parameters of the optics}
 \begin{tabular}{lr}
  \hline
  Parameter & Value [mm]\\
  \hline
  \underline{Parameters of the main and the sub-ref.} & \\
  Diameter of main-ref. & 45,000.00\\
  Focal length of main-ref. & 16,000.00\\ 
  Diameter of sub-ref. & 3,750.00\\
  Interfocal distance of primary focus&\\
  and secondary focus & 21,780.60\\
  &\\
  \underline{Distance} & \\
  Sub-ref. to M1 (plane) & 20,211.90\\
  M1 to M2 (ellipsoidal) & 65,006.00\\ 
  M2 to M3 (ellipsoidal) & 17,000.00\\ 
  M3 to M4 (plane) & 2,300.00\\
  M4 to M55 (plane) & 3,000.00\\ 
  M55 toM56 (ellipsoidal) & 2,500.00\\ 
  M56 to M59 (plane) & 1,200.00\\ 
  M59 to M60 (ellipsoidal) & 1,200.00\\
  M60 to EM (ellipsoidal) & 2,300.00\\
  EM to corrugated horn & 200.00\\
  &\\
  \underline{Focal distances of the ellipsoidal mirrors} & \\
  M2 & 6,296.00\\
  M3 & 6,296.00\\
  M56 & 1,633.00\\
  M60 & 1,633.00\\
  EM & 1,847.00\\
  &\\
  \underline{Parameters of the horn} & \\
  Diameter of horn aperture & 36.67\\
  Slantlength of horn & 115.19\\
\hline
 \end{tabular}
\end{center}
\end{table}

\section{Introduction}
The 45-m telescope is located at Nobeyama Radio Observatory (NRO)\footnote{The Nobeyama Radio Observatory is a branch of the National Astronomical Observatory of Japan, National Institutes of Natural Sciences. (http://www.nro.nao.ac.jp/index-e.html)} in Nagano, Japan and is one of the largest millimeter-wave telescopes in the world (e.g., Morimoto 1981; Akabane 1983). The 45-m telescope is equipped with a low-noise High Electron Mobility Transistor (HEMT) and Superconductor-Insulator-Superconductor (SIS) receivers covering the range of 20 to 115 GHz, along with powerful spectral-line and continuum back-ends. Scientific discoveries using this instrument include the discovery of a super massive black hole (Nakai et al.\ 1993; Miyoshi et al.\ 1995) as well as the discovery of a number of interstellar molecules (e.g., Kawaguchi et al.\ 1995; Kaifu et al.\ 2004) and cover wide range of research fields, such as the formation of stars and planetary systems (e.g., Tatematsu et al.\ 1993; Skrutskie et al.\ 1993; Mizuno et al.\ 1994), the structure and activity of galaxies (e.g., Nakai et al.\ 1994; Kuno et al.\ 2007), and interstellar chemistry (e.g., Suzuki et al.\ 1992; Hirahara et al.\ 1992; Takano et al.\ 1998; Sakai et al.\ 2006).
The 100-GHz band SIS receivers are the most important receivers for the 45-m telescope, because they cover the highest frequency range among the receivers installed in the telescope. From a scientific viewpoint, the $J$ = 1--0 emission line of carbon monoxide (CO) in this frequency band is a principal probe for studies of interstellar molecular gas. The SIS-80 (S80) and SIS-100 (S100) single beam receivers, which cover radio frequency (RF) bands of 72--115 GHz and 77--115 GHz respectively, are installed in the 45-m telescope. However, as the receiver noise temperature decreases throughout the world (e.g., Shi et al.\ 1997; Chin et al.\ 2004; Asayama et al.\ 2004), the performances of S80 and S100 have relatively worsened since their installation. Moreover, the most important cause of high noise temperature is that these receivers use a wire grid for the separation of polarizations and a Martin-Puplett Interferometer (MPI) as an image rejection filter for the separation of sidebands. The loss with the quasi-optics results in the degradation of the system noise temperature. These quasi-optics are too sensitive for their optical alignment, and thus misalignment becomes another cause of degradation. The Single Side-Band (SSB) receiver noise temperatures including the quasi-optics are 150--300 K for both receivers.
To solve these problems, we began the development of a new receiver system. The new receiver system uses a waveguide-type Ortho-Mode Transducer (OMT) and two Sideband-Separating (2SB) SIS mixers. To date, simultaneous detection of molecular emission lines with a waveguide-type 2SB receiver has been performed by a few groups toward a limited number of directions in the sky (e.g., Asayama et al.\ 2003a; Lauria et al.\ 2006), suggesting that the 2SB receiver systems are more efficient compared to other SSB receivers (e.g., Nakajima et al.\ 2007). As a result, we can reduce the number of the optical elements and the composition becomes simple. In addition, it is also noteworthy that through the simultaneous detection of two polarizations and two sidebands using a single optical horn, we can better determine the line intensity ratios of some molecular lines without being contaminated by errors in pointing.

We developed a new 2SB receiver system in the 100-GHz band for the 45-m telescope. Over the RF range of 80--120 GHz, the SSB receiver noise temperature of the mixer is lower than 100 K for the 4.0--8.0 GHz Intermediate Frequency (IF) band. The Image Rejection Ratios (IRRs) are greater than 10 dB over the same range. It is confirmed that the typical SSB receiver noise temperature of the new system is approximately half that of the previous systems of S80 and S100. The IF bandwidth of the new receiver system is 4 GHz for each IF output, while those of previous receivers are 600 MHz. We can detect 16 GHz in total. Using the newly developed 100-GHz band SIS receiver system for 45-m telescope, we observed the $^{12}$CO, $^{13}$CO and C$^{18}$O emission lines simultaneously toward the Sagittarius B2 region in order to confirm the performance of the receiver system. This is the first astronomical observation using a waveguide-type dual-polarization sideband-separating SIS receiver system in the 100-GHz band. The SSB noise temperatures of the system, including the atmosphere, are approximately 180 K in the LSB at $f_{\mathrm{LO}}$ = 109 GHz ($f_{\mathrm{RF}}$ = 103 GHz) for pol-1 and at $f_{\mathrm{LO}}$ = 104 GHz ($f_{\mathrm{RF}}$ = 98 GHz) for pol-2 at an elevation of 80$^\circ$. The system noise temperature, including the atmosphere, became approximately half that of the previous receiver system (S100). The IRRs of the two 2SB mixers were calculated from the $^{12}$CO spectra at 115.271 GHz from the W51 giant molecular cloud, resulting in $>$ 20 dB for one polarization and $>$ 14 dB for the other polarization.

In the present paper, we describe the 2SB receiver system in the 100-GHz band for 45-m radio telescope and demonstrate its performance. The results of the test observation using the dual-polarization and sideband-separating SIS receiver for the 100 GHz band are also presented.

\begin{figure}
  \begin{center}
  \FigureFile(80mm,52mm){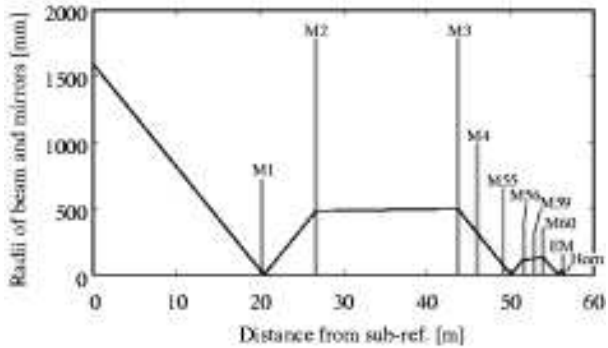}
  \end{center}
  \caption{Gaussian beam propagation at 100 GHz. To reduce cross polarization loss, M2 and M3, and M56 and M60 have the same shapes, respectively. The lengths of the vertical bars denote the radii of the mirrors.}\label{fig2}
\end{figure}

\begin{table*}
\begin{center}
\caption{RF parameters of the optics}
 \begin{tabular}{lrrr}
  \hline
  Parameter & & Value [mm]&\\
  \hline
  & \multicolumn{3}{c}{Frequency [GHz]}\\ \cline{2-4}
  & 84 & 100 & 116\\
  \hline
  Beam size at sub-ref. & 1,595.21 & 1,595.21 & 1,595.21\\
  Curvature at sub-ref. & 20,489.17 & 20,489.17 & 20,489.17\\ 
  Beam waist between sub-ref. and M2 & 14.59 & 12.26 & 10.57\\
  Beam size at M2 & 484.86 & 484.75 & 484.68\\
  Curvature at M2 & 6,230.16 & 6,227.99 & 6,226.66\\ 
  Curvature at M2 (image) & $-$595,725.99 & $-$576,560.98 & $-$565,356.75\\ 
  Beam waist between M2 and M3 & 458.01 & 445.79 & 432.63\\ 
  Beam size at M3 & 500.28 & 500.16 & 500.09\\ 
  Curvature at M3 & 501,358.69 & 515,317.62 & 524,310.71\\ 
  Curvature at M3 (image) & 6,376.07 & 6,373.87 & 6,372.52\\
  Beam waist between M3 and M56 & 14.47 & 12.16 & 10.48\\
  Beam size at M56 & 113.12 & 112.89 & 112.75\\
  Curvature at M56 & 1,453.05 & 1,446.67 & 1,442.74\\
  Curvature at M56 (image) & $-$13,186.05 & $-$12,678.45 & $-$12,383.11\\
  Beam waist between M56 and M60 & 86.01 & 77.72 & 70.50\\
  Beam size at M60 & 135.86 & 135.79 & 135.74\\
  Curvature at M60 & 13,288.42 & 13,488.11 & 13,614.27\\
  Curvature at M60 (image) & 1,861.79 & 1,857.94 & 1,855.57\\
  Beam waist between M60 and EM & 15.47 & 13.00 & 11.21\\
  Beam size at EM & 37.32 & 36.13 & 35.37\\
  Curvature at EM & 558.23 & 527.33 & 508.07\\ 
  Curvature at EM (image) & 274.47 & 282.61 & 288.47\\
  Beam waist between EM and horn & 8.15 & 7.31 & 6.59\\
  Beam size at horn aperture & 11.81 & 11.81 & 11.80\\
  Curvature at horn aperture & 117.18 & 115.19 & 114.00\\
  \hline
 \end{tabular}
\end{center}
\end{table*}

\section{Optics}
The 45-m telescope has a Nasmyth optical system. A signal from space focused by a Cassegrain system is reflected by a plane mirror (M1) so as to pass along the elevation axis and is led to a receiver cabin by a pair of ellipsoidal mirrors (M2 and M3). There are several receivers in the receiver cabin. By changing the arrangement of the mirrors (e.g., M4), a signal can be led to a desired receiver. In order to install the new receiver system, we designed a receiver optical system. The newly developed optical elements include a corrugated horn and an ellipsoidal mirror (figure 1).

\begin{figure}
  \begin{center}
  \FigureFile(80mm,64mm){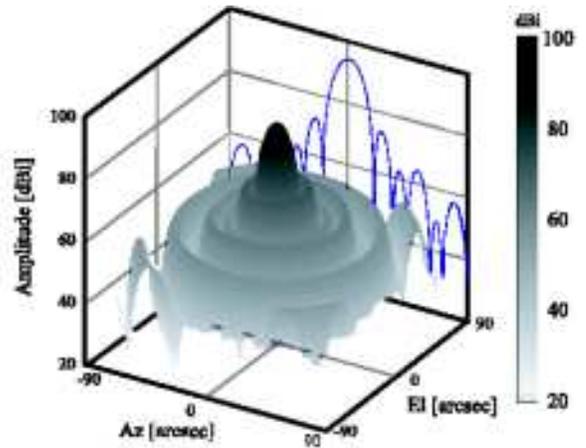}
  \end{center}
  \caption{Antenna beam pattern at 100 GHz as calculated by GRASP 9 physical optics software.}\label{fig3}
\end{figure}

We used the method of Gaussian optics and physical optics to design the optics. We first designed the optics in Gaussian optics (Goldsmith 1998) because Gaussian optics is generally used for optical design in the millimeter wave band. Assuming the edge taper level of the sub-reflector as $-$12 dB, we designed the new optics based on a Gaussian beam propagation. Figure 2 shows the beam propagation at 100 GHz, and the optical parameters are listed in tables 1 and 2. 

\begin{table*}
\begin{center}
\caption{Physical dimensions of the corrugated horn}
 \begin{tabular}{lc}
 \hline
  parameters & value\\
 \hline
  Material & Copper\\
  Total length & 105.00 mm\\
  Length of the corrugation section & 102.60 mm\\
  Outer diameter of the aperture & 42.00 mm\\
  Inner diameter of the aperture & 36.70 mm\\
  Semi-flare angle & 9.3$^\circ$\\
  Diameter of the circular waveguide & 3.23 mm\\
  Width of the grooves & 0.54 mm\\
  Depth of the grooves at the mode-launching section & 1.28--0.88 mm\\
  Depth of the grooves at the thread section & 0.83 mm\\
 \hline
 \end{tabular}
\end{center}
\end{table*}

We designed the new ellipsoidal mirror with an edge level of $-$54 dB so that the spillover loss would be small. In this design, more than 99\% of the total power can be transmitted. We fabricated the ellipsoidal mirror using an NC milling cutter and checked the surface error using a 3D vision measurement system. As a result, a surface error of 3.8 $\mu$m r.m.s.\ is achieved, which is sufficiently small at the observation wavelength of $\sim$3 mm. This mirror is fixed on the side of a receiver dewar by a holder having a positional adjustment system.

\begin{figure}
  \begin{center}
  \FigureFile(80mm,31mm){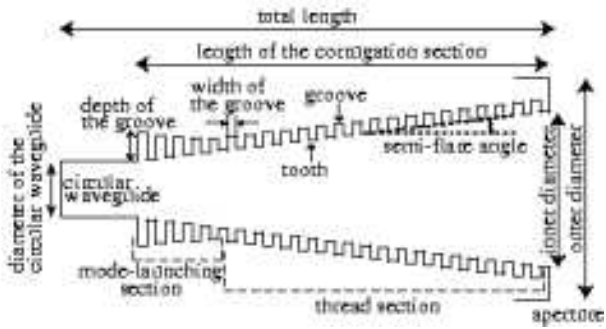}
  \end{center}
  \caption{Cutaway view of the schematic diagram of corrugated horn.}\label{fig4}
\end{figure}

\begin{figure}
  \begin{center}
  \FigureFile(80mm,65mm){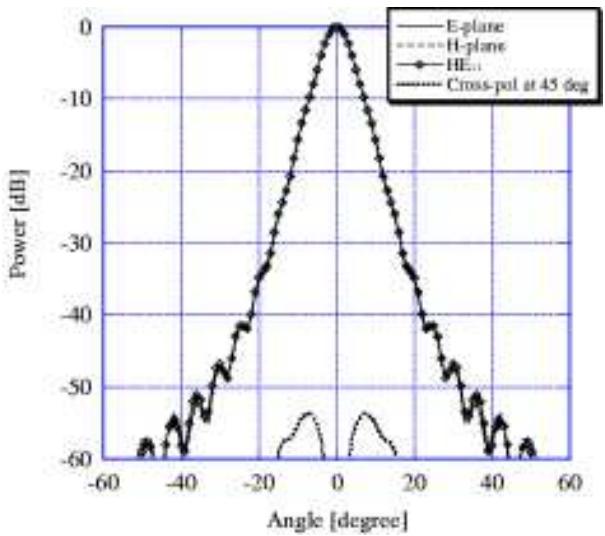}
  \end{center}
  \caption{Simulated beam profile of the corrugated horn at 100 GHz. The result shows the similarity of the beam patterns between the E-plane and the H-plane down to $-$50 dB. These beam patterns correspond well with the theoretical beam pattern obtained by the radiation of the HE$_{11}$ mode.}\label{fig4}
\end{figure}

Next, we used the technique of physical optics for the evaluation of the optical system obtained by Gaussian optics. In this calculation, we used the GRASP9 analysis software package of the reflector antenna (TICRA). As a result, an antenna directivity of 92.15 dBi, which is equivalent to an aperture efficiency $\eta$ of 0.74 for the 45-m telescope, a beam size of 15.8 arcsec, and the first side lobe level of $-$17 dB were obtained (figure 3). However, the actual performance may be somewhat worse by several percent because the calculation was made without blocking by the sub-reflector stays, surface error, or ohmic loss of the optical elements. 

Corrugated horns are commonly used with reflector antenna systems. Since corrugated horns can reduce the edge diffraction, improved pattern symmetry and reduced cross polarization can be obtained (Clarricoats et al.\ 1984). Such performance is needed for the horn used for the wide range of the RF frequency. We repeatedly calculated the suitable corrugation pattern from the basic design. The physical dimensions of the corrugated horn are shown in table 3 and figure 4. As a result, we obtained a return loss lower than $-$23 dB, a maximum cross polarization level lower than $-$30 dB, and good similarity between the calculated beam profiles of the E-plane and H-plane (figure 5). We fabricated the horn by the directly-dig method, rather than by the electro-forming method (Kimura et al.\ 2008).

\section{Receiver}
\subsection{Receiver configuration}
The specifications of the receiver system are summarized in table 4 and a block diagram of the receiver system is shown in figure 6. The operation frequency range is from 72 to 128 GHz, which is limited by the frequency range of the LO chain. The IF frequency range of 4.0--8.0 GHz with the bandwidth of 4.0 GHz is adopted. This receiver is composed of an OMT and two 2SB mixers, which are both based on a waveguide technique. The 2SB mixers are hereafter referred to as mixer-A and mixer-B. The photograph of the receiver in the dewar is shown in figure 7.

\begin{figure*}
  \begin{center}
  \FigureFile(160mm,118mm){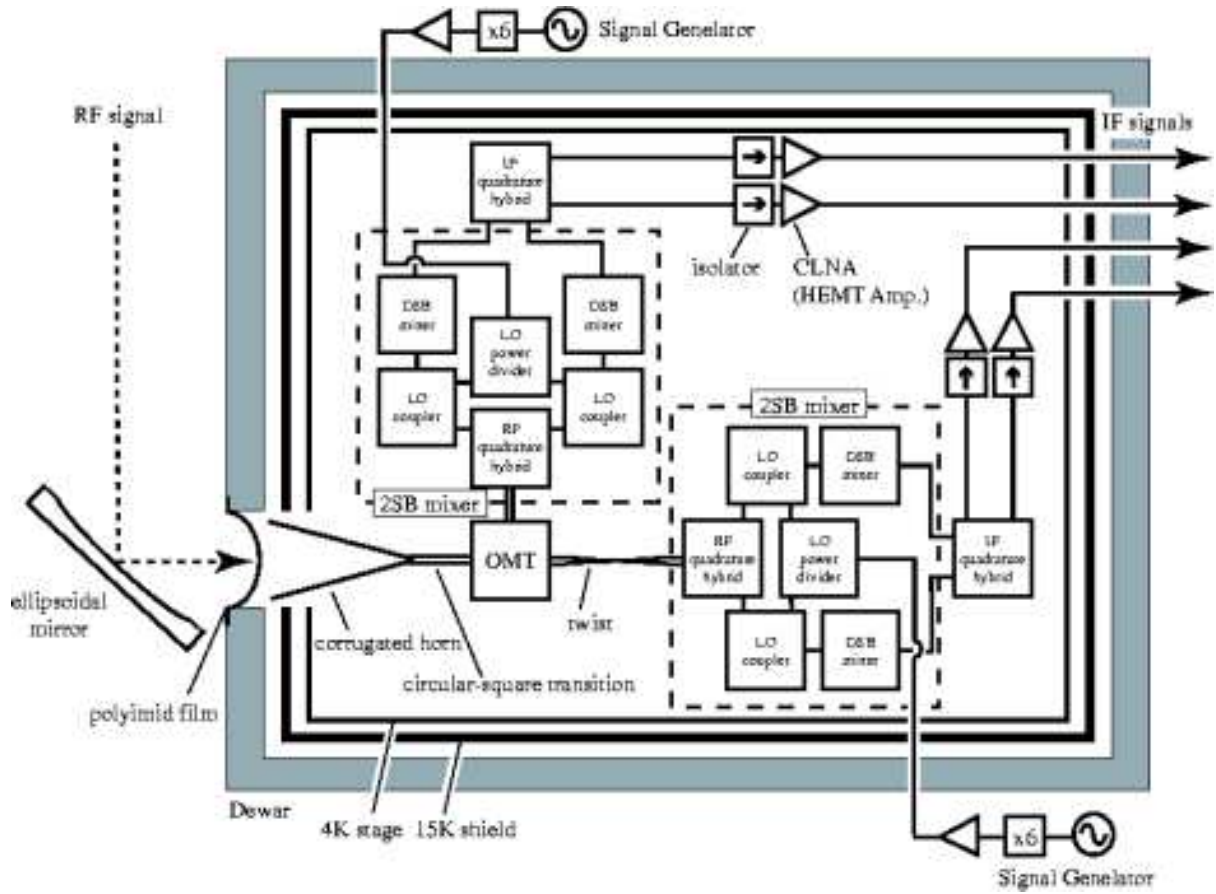}
  \end{center}
  \caption{Block diagram of the dual-polarization sideband-separating receiver system. Two orthogonal polarizations of the RF signal into a single horn are separated by an OMT, and two sidebands of each polarization are separated by two 2SB mixers. Thus, we can obtain four IF signals independently and simultaneously.}\label{fig6}
\end{figure*}

\begin{figure}
  \begin{center}
  \FigureFile(80mm,60mm){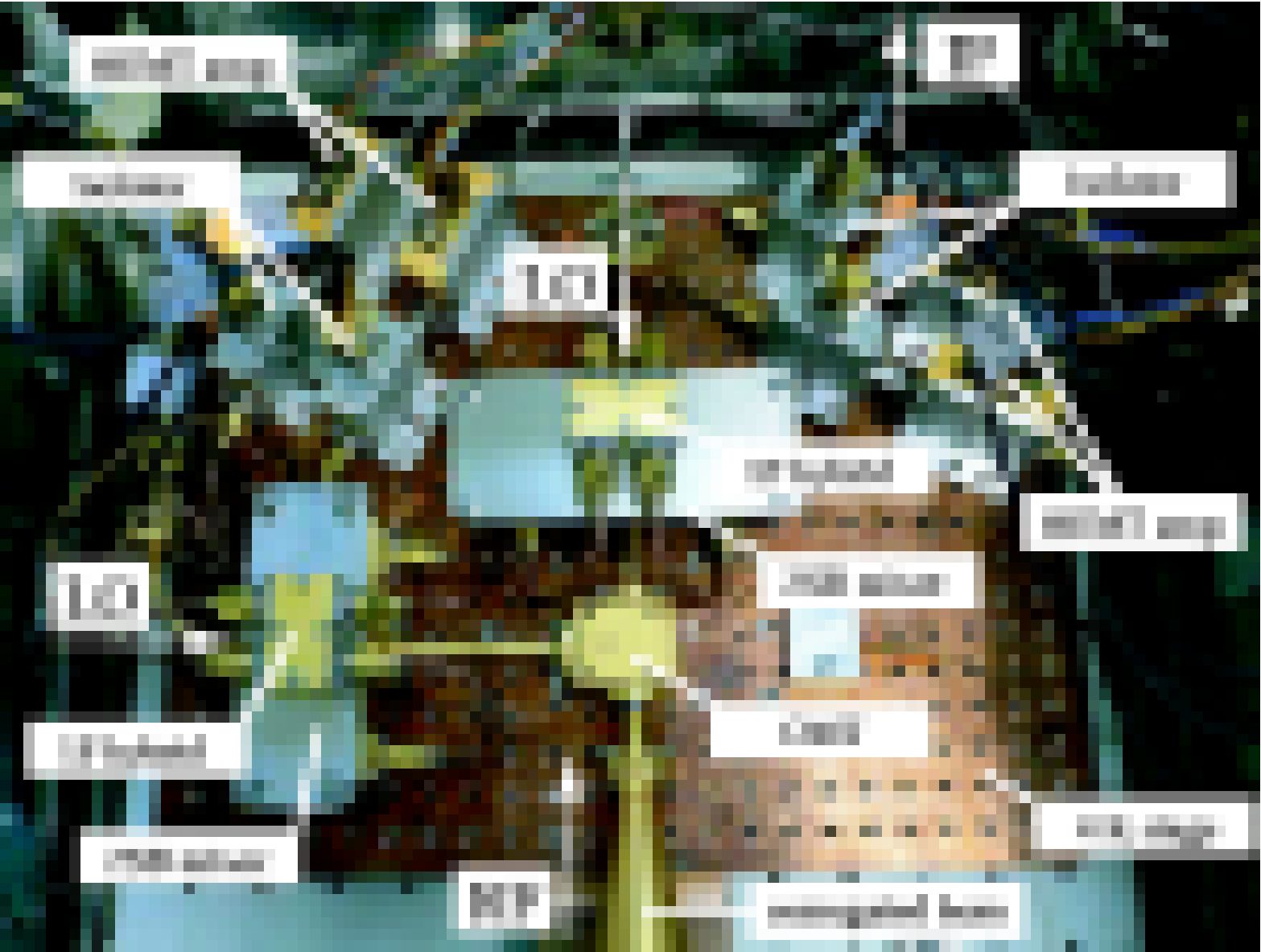}
  \end{center}
  \caption{Photograph of the 4 K cooled stage in the receiver dewar. The RF signal is fed into the OMT located in the middle of the figure from the lower side using a detachable corrugated horn. The LO signals are fed from the upper side for the 2SB mixer located above the OMT and from the left side for the other 2SB mixer.}\label{fig7}
\end{figure}

By using an OMT, it is possible to receive signals in the dual linear polarizations independently and simultaneously. The OMT adopted herein was developed at the Australia Telescope National Facility (ATNF), Commonwealth Scientific and Industrial Research Organization (CSIRO), and consists of a square to double ridged guide transition followed by a junction of two side arms with the central guide (Moorey et al.\ 2006). Measurement of the OMT performance by ATNF has revealed a return loss of $>$ 20 dB from 71 to 118 GHz and an insertion loss at room temperature of $<$ 0.16 dB for both polarizations.

\begin{figure}
  \begin{center}
  \FigureFile(80mm,80mm){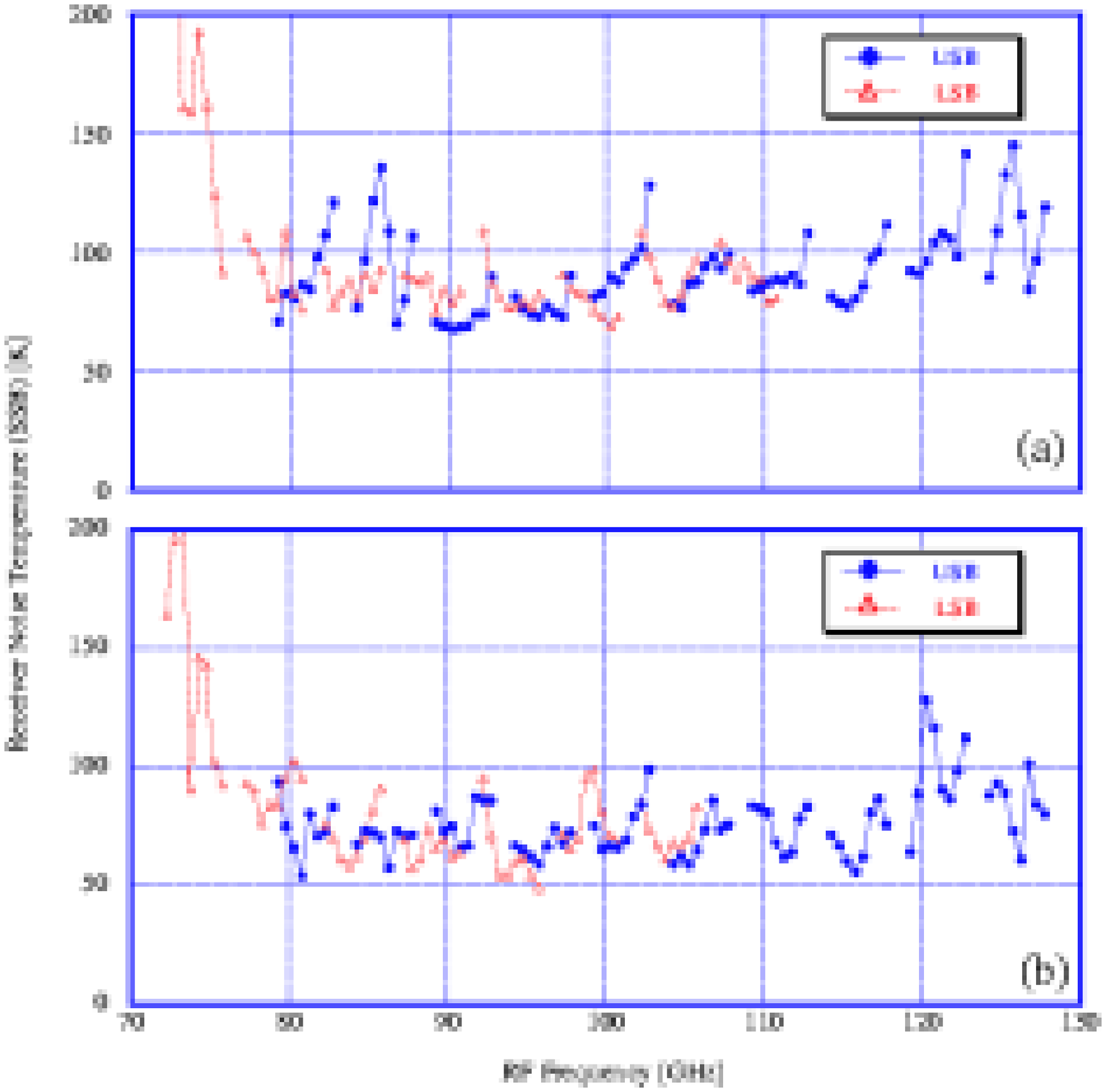}
  \end{center}
  \caption{SSB receiver noise temperatures in the LSB and USB of (a) mixer-A and (b) mixer-B. These values are measured by 4.0-8.0 GHz IF and are shown as a function of RF frequency (see text).}\label{fig8}
\end{figure}

The detailed structure of a split-block waveguide unit for our 2SB mixer is described in Asayama et al.\ (2004). The basic design of the present sideband-separating SIS mixer is similar to that described in Claude et al.\ (2000). The split-block waveguide unit contains an RF quadrature hybrid, two LO directional couplers, an LO power divider, and 4 K cold image terminations. We also integrated two DSB mixers on the split-block waveguide unit through the waveguide taper transformers. The RF and LO signals are fed to the feed point through a linearly tapered waveguide impedance transformer, which uses a full height to 1/5 reduced height waveguide for waveguide-to-stripline transition of the SIS mixer. The linearly tapered waveguide impedance transformer was designed using the lumped-gap-source port provided by HFSS$^{\rm TM}$ (Asayama et al.\ 2003b). The SIS junctions adopted herein were developed at the NRO. A six-series array was composed of Nb/AlO$_{x}$/Nb junctions with a normal state resistance of approximately 80--90 $\Omega$, and each junction area was 1.9 $\times$ 1.9 $\mu$m$^{2}$. The reason for using the series junction is that a wider bandwidth of RF frequency can be achieved. In the present system, the required fractional bandwidth totals ${\gtrsim}$ 40\%. The Parallel Connected Twin Junction (PCTJ) designed by Asayama et al.\ (2004) is not suited for this receiver because the designed RF frequency range is 90--115 GHz, which corresponds to a fractional bandwidth of only approximately 25\%. Moreover, the series junction barely saturates and the intensity can be calibrated with high accuracy (Kerr 2002). The IF signals from the two DSB mixers are combined in a commercial quadrature hybrid (Nihon Tsushinki Inc.).

\begin{table*}
\begin{center}
\caption{Specifications of the receiver system}
 \begin{tabular}{ll}
  \hline
  Tuning range & 72--128 GHz\\
  Mixer type & SIS (Nb/AlO$_{x}$/Nb) six-series junction\\
  & mounted on fixed-tuned waveguide-type 2SB mixer block\\
  Receiving mode & SSB (2SB) operation\\
  Polarization & dual linear polarization\\
  Receiver noise temperature (DSB) & $\sim$ 50 K ($f_{\mathrm{LO}}$ = 80--115 GHz)\\
  & \\
  \underline{mixer-A} & \\
  Receiver noise temperature (SSB) & ${\lesssim}$ 100 K ($f_{\mathrm{RF}}$ = 75--120 GHz)\\
  & $\sim$ 68 K (minimum value at $f_{\mathrm{RF}}$ = 90 GHz in the USB)\\
  & LSB : 86.8 K (average value at $f_{\mathrm{RF}}$ = 75--106 GHz)\\
  & USB : 88.2 K (average value at $f_{\mathrm{RF}}$ = 79--120 GHz)\\
  Image rejection ratio & $>$ 10 dB ($f_{\mathrm{RF}}$ = 80--123 GHz)\\
  & LSB : 14.9 dB (average value at $f_{\mathrm{RF}}$ = 80--111 GHz)\\
  & USB : 14.3 dB (average value at $f_{\mathrm{RF}}$ = 84--123 GHz)\\
  & \\
  \underline{mixer-B} & \\
  Receiver noise temperature (SSB) & ${\lesssim}$ 100 K ($f_{\mathrm{RF}}$ = 75--120 GHz)\\
  & $\sim$ 49 K (minimum value at $f_{\mathrm{RF}}$ = 95 GHz in the LSB)\\
  & LSB : 74.5 K (average value at $f_{\mathrm{RF}}$ = 75--106 GHz)\\
  & USB : 72.7 K (average value at $f_{\mathrm{RF}}$ = 79--120 GHz)\\
  Image rejection ratio & $>$ 10 dB ($f_{\mathrm{RF}}$ = 80--100 GHz)\\
  & LSB : 13.4 dB (average value at $f_{\mathrm{RF}}$ = 80--100 GHz)\\
  & USB : 12.2 dB (average value at $f_{\mathrm{RF}}$ = 84--110 GHz)\\
  & \\
  IF frequency band & 4.0--8.0 GHz (4.0 GHz bandwidth)\\
  IF amplifier & cooled low noise HEMT amplifier\\
  & (typical noise temperature of 8 K and gain of $+$30 dB)\\
  \hline
  \multicolumn{2}{@{}l@{}}{\hbox to 0pt{\parbox{85mm}\hss}}
 \end{tabular}
\end{center}
\end{table*}

\subsection{Noise temperature}
Before installing the receiver system in the telescope, we evaluated its performance. The noise temperature of the 2SB SIS receiver was measured by a standard Y-factor method using hot (300 K) and cold (77 K) loads in the laboratory. The mixer was mounted on a 4 K cold stage in a dewar. The first-stage IF amplifier is a 4 K cooled HEMT at the 4.0--8.0 GHz band. The equivalent noise temperature and the gain of the HEMT amplifier associated with an isolator were approximately 8 K and +30 dB, respectively. The following-stage amplifiers work at room temperature. 
The measured DSB receiver noise temperatures of the SIS mixers by 4.0--8.0 GHz IF by a power meter are approximately 50 K over the LO frequency range of 80--115 GHz. The SSB receiver noise temperature, including the noise contribution from the vacuum window, the feed horn, and the IF amplifier chain, were measured by a spectrum analyzer. For a certain LO frequency, the output spectra were smoothed to a resolution of 500 MHz and sampled at eight IF frequency points from 4.25 GHz to 7.75 GHz at intervals of 0.5 GHz. We then swept the LO frequency from 80 GHz to 120 GHz by a step of 5 GHz to cover the RF frequency range from 72.25--127.75 GHz. Finally, the measured SSB receiver noise temperatures were corrected for the IRR, in order to obtain the SSB receiver noise temperature. Since the IRRs were measured with the different measurement system from the one used to measure the noise temperature, we assumed the IRRs to be 10 dB, that is nearly equal to the smallest value among the actual IRRs (see section 3.3), for all the measured points. Therefore, it is to be noted that the actual SSB receiver noise temperature will be better than the values presented in this paper for most of the measured points. The SSB receiver noise temperatures were measured to be lower than approximately 100 K over the RF range of 75--120 GHz, and the minimum value of $\sim$68 K for mixer-A in the USB and $\sim$49 K for mixer-B in the LSB are achieved at approximately 90 GHz and 95 GHz, respectively. The mean value and standard deviation between the RF frequency range of 75--106 GHz in the LSB and 79--120 GHz in the USB are 87$\pm$11 K and 88$\pm$15 K, respectively, for mixer-A (figure 8(a)) and are 74$\pm$14 K and 73$\pm$10 K, respectively, for mixer-B (figure 8(b)). 

\begin{figure}
  \begin{center}
  \FigureFile(80mm,80mm){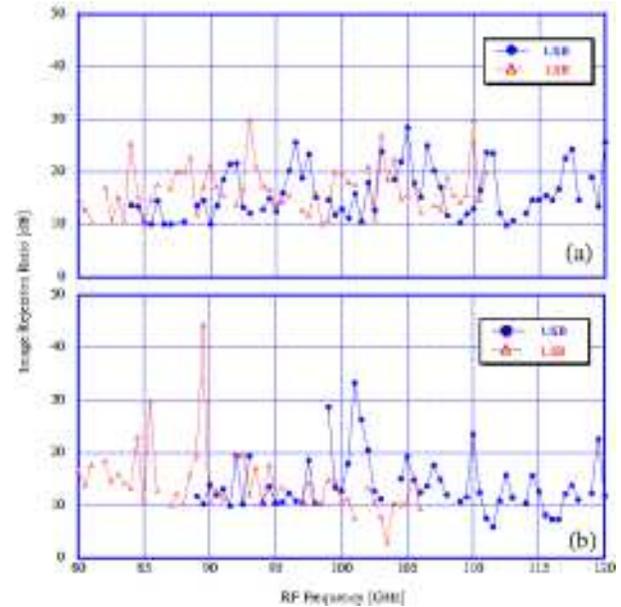}
  \end{center}
  \caption{Image rejection ratio in the LSB and USB of (a) mixer-A and (b) mixer-B. These values are measured by 4.0-8.0 GHz IF and are shown as a function of RF frequency.}\label{fig9}
\end{figure}

\subsection{Image rejection ratio}
The IRRs were measured by the relative amplitudes of the IF responses in the USB and LSB when injecting a Continuous Wave (CW) signal (Kerr et al.\ 2001) from a signal generator. The IRR measured by 4.0--8.0 GHz IF was greater than 10 dB over the RF frequency range of 80--123 GHz for mixer-A (figure 9(a)). The mean value between the RF frequency range of 80--111 GHz in the LSB and 84--123 GHz in the USB are 15 dB and 14 dB, respectively. The IRR of mixer-B was lower than 10 dB at $f_{\mathrm{LO}} {\gtrsim} $105 GHz (figure 9(b)). However, both the mean value between the RF frequency range of 80--102 GHz in the LSB and that between the RF frequency range of 84--110 GHz in the USB were 12 dB. Therefore, it is most desirable to use mixer-B for observation at lower RF frequency. There is a possibility in the future that the mixer will be replaced by a new mixer with better performance.

\begin{figure}
  \begin{center}
  \FigureFile(80mm,107mm){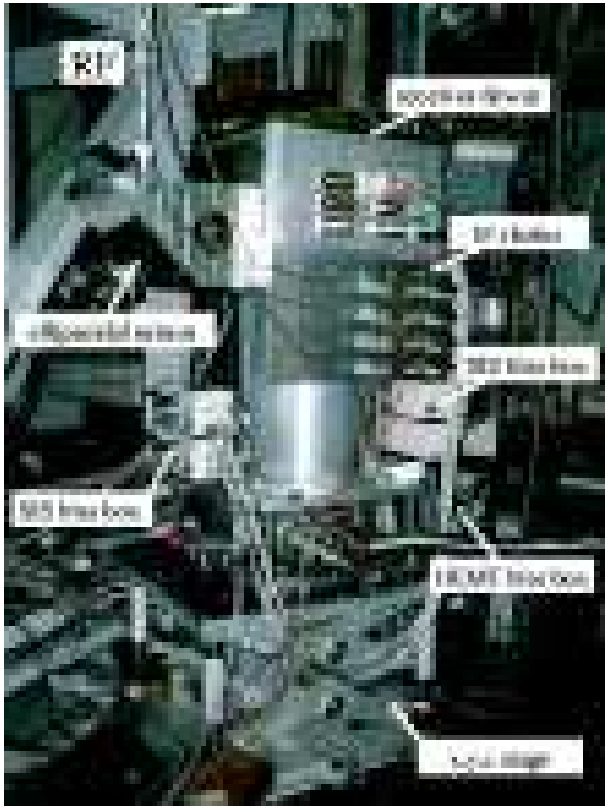}
  \end{center}
  \caption{Photograph of the receiver system in the receiver cabin of the telescope.}\label{fig10}
\end{figure}

\subsection{Installation}
The receiver system was installed in the 45-m telescope in early December 2007. Figure 10 shows the receiver dewar in the receiver cabin. The SIS and HEMT amplifier biases, LO oscillators, and IF chains are placed around the dewar. The LO signals for each 2SB mixer of dual-polarization are independently generated by multiplying the output of the signal generators ($\sim$12--20 GHz) with 2$\times$3 multipliers. Therefore, the four IF signals from two 2SB mixers can cover different regions of the RF frequency band. Hereafter, the polarization for mixer-A is referred to as pol-1, and the polarization for mixer-B is referred to as pol-2.

\section{Results}
\subsection{System noise temperature}
The performance of the receiver may change when it is installed in the telescope because the environment of the receiver is different from that in the laboratory. Therefore, we installed the receiver system in the telescope and measured the performance of the receiver system. Note that the noise temperatures in this subsection are not corrected for IRR. Even after correcting for IRR, the noise temperature increases only ${\lesssim}$ 10\%. The SSB receiver noise temperature, including the ellipsoidal mirror before the horn, is measured by a standard Y-factor method to be approximately 60 K in the USB at $f_{\mathrm{LO}}$ = 109 GHz ($f_{\mathrm{RF}}$ = 115 GHz) for both 2SB mixers. The SSB noise temperatures of the system, including the atmosphere, are approximately 180 K in the LSB at $f_{\mathrm{LO}}$ = 109 GHz ($f_{\mathrm{RF}}$ = 103 GHz) for pol-1 and at $f_{\mathrm{LO}}$ = 104 GHz ($f_{\mathrm{RF}}$ = 98 GHz) for pol-2 at an elevation of 80$^\circ$. The system noise temperature, including the atmosphere, became approximately half of that of the previous receiver system (S100).

\subsection{Beam size and main-beam efficiency}
We estimated the beam size and the main-beam efficiency of the 45-m telescope based on observations of the quasar, 3C273, and the Saturn in continuum. Note that the following data is obtained at the beginning of the commissioning of the new receiver system, and thus the results are preliminary. We scanned 3C273 in the azimuth and elevation directions, while recording the total power of the IF output. The beam pattern of the antenna is obtained from the resultant map of 3C273 by rotating 180$^\circ$ along the peak position, because 3C273 can be assumed as a point source. Figure 11 shows the resultant map of 3C273 at 115 GHz. The half power beam width (HPBW) of the telescope is estimated to be 15.$\!''$5$\pm$0.1 for pol-1 and 17.$\!''$6$\pm$0.3 for pol-2 assuming that the beam pattern has a Gaussian shape in both polarizations. The main-beam efficiency, $\eta_{\mathrm{mb}}$, was calculated to be 34 \% and 35 \% for pol-1 and pol-2, respectively, assuming that the brightness distribution of Saturn as a uniform disk of apparent diameter of 18$\!''$ with a brightness temperature of 150 K (Mangum 1993). 

\begin{figure}
  \begin{center}
  \FigureFile(80mm,40mm){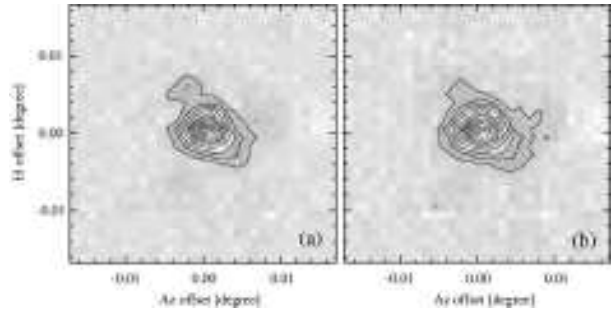}
  \end{center}
  \caption{Resultant map of 3C273 at 115 GHz. Contour intervals and the lowest contour are both 10\% of the maximum value. The output of the 2SB mixer for pol-1 and pol-2 are shown in (a) and (b), respectively. A similar beam pattern is confirmed in both polarizations.}\label{fig11}
\end{figure}

\subsection{Test observation}
The first astronomical signal after the installation, $^{12}$CO ($J$ = 1--0) spectra at 115.271 GHz in the USB from the W51 giant molecular cloud, was obtained on December 11, 2007 (figure 12(a) and (b)). We also obtained HCO$^{+}$ ($J$ = 1--0) spectra at 89.182 GHz in the LSB from the W51 on December 17, 2007 (figure 12(c) and (d)). The IRRs of the two 2SB mixers were calculated from these spectra, resulting in $>$ 20 dB in both the USB at $f_{\mathrm{RF}}$ $\sim$115 GHz and the LSB at $f_{\mathrm{RF}}$ $\sim$89 GHz for pol-1 (figure 12 (a) and (c)) and $\sim$14 dB in the USB at $f_{\mathrm{RF}}$ $\sim$115 GHz and $\sim$12 dB in the LSB at $f_{\mathrm{RF}}$ $\sim$89 GHz for pol-2 (figure 12 (b) and (d)). These values are as good as those obtained in the laboratory. Moreover, we successfully observed the molecular lines toward the Sagittarius B2 region by using six digital spectrometers (Sorai et al.\ 2000) with bandwidths of 512 MHz. Figure 13 shows an example of the results, which are obtained by a single pointing. We detected CH$_{3}$C$_{2}$H and H$_{2}$CS in the LSB and $^{12}$CO in the USB, respectively, for pol-1, and CS in the LSB and SO$_{2}$, C$^{18}$O, HNCO, $^{13}$CO and CH$_{3}$CN in the USB, respectively, for pol-2. Since four IF signals can be observed independently and simultaneously, we obtained these spectra by integrating the signal only for five minutes. This was the first astronomical observation using the waveguide-type dual-polarization sideband-separating SIS receiver system in the 100 GHz band.

\begin{figure}
  \begin{center}
  \FigureFile(80mm,80mm){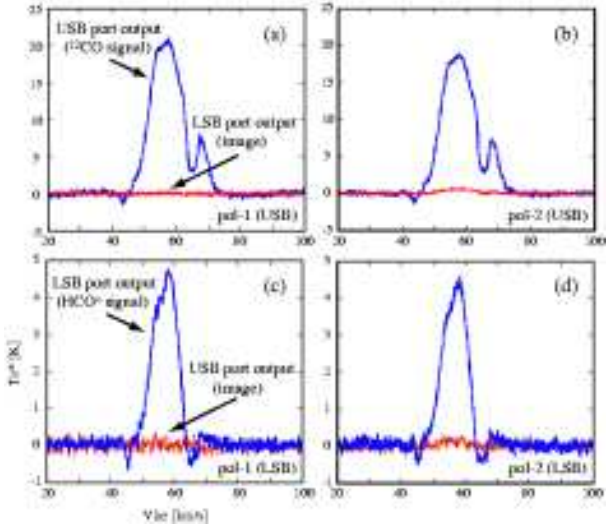}
  \end{center}
  \caption{IF signal outputs of the USB and LSB ports obtained by observations toward W51. The outputs of the 2SB mixer for pol-1 are shown in (a) and (b), and that for pol-2 are shown in (c) and (d). The signal of the $^{12}$CO line in the USB port (bold line) leaks to the LSB port (thin line) in panels (a) and (b), and the signal of the HCO$^{+}$ line in the LSB port (bold line) leaks to the USB port (thin line) in panels (c) and (d). Note that the absorption at $V_{\mathrm{lsr}}$ $\sim$45 km s$^{-1}$ and $\sim$65 km s$^{-1}$ may be caused by the emission in the OFF position.}\label{fig12}
\end{figure}

\begin{figure}
  \begin{center}
  \FigureFile(80mm,135mm){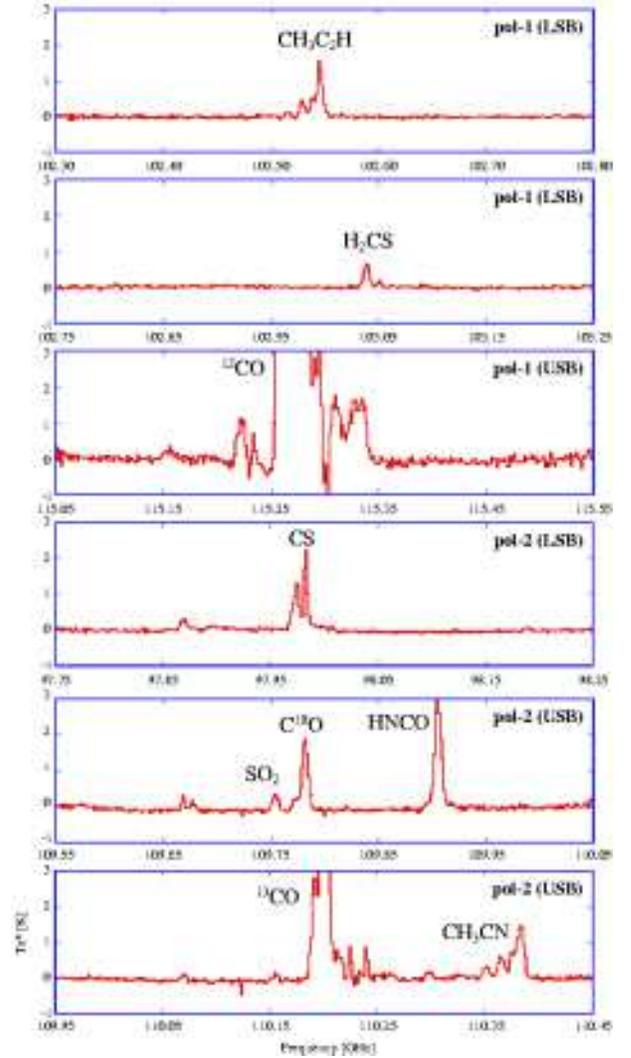}
  \end{center}
  \caption{Results of the line survey toward Sgr B2. The four IF signals were obtained independently and simultaneously by a single pointing.}\label{fig13}
\end{figure}

\section{Conclusions}
A waveguide-type dual-polarization sideband-separating SIS receiver system in the 100 GHz band was developed for and installed in the NRO 45-m telescope. This receiver system is composed of an ortho-mode transducer and two sideband-separating SIS mixers, both of which are based on the waveguide type. The four IF bands, 4.0--8.0 GHz, in each of the polarizations and sidebands can be observed simultaneously and separately. Over the RF frequency range of 80--120 GHz, the single-sideband receiver noise temperatures are 50--100 K and the image rejection ratios are greater than 10 dB. We have developed new optics for the receiver and new IF chains for the four IF signals obtained by the receiver system. The new receiver system was installed in the 45-m telescope, and we successfully observed the $^{12}$CO, $^{13}$CO, and C$^{18}$O emission lines simultaneously toward the Sagittarius B2 region. This is the first astronomical observation using a waveguide-type dual-polarization sideband-separating SIS receiver system in the 100 GHz band. The SSB noise temperatures of the system, including the atmosphere, are approximately 180 K in the LSB at $f_{\mathrm{LO}}$ = 109 GHz ($f_{\mathrm{RF}}$ = 103 GHz) for one polarization, and at $f_{\mathrm{LO}}$ = 104 GHz ($f_{\mathrm{RF}}$ = 98 GHz) for the other polarization at an elevation of 80$^\circ$. The SSB system noise temperature, are measured to be $\sim$180 K at $f_{\mathrm{RF}}$ $\sim$100 GHz for both polarizations, and therefore became approximately half of that of the previous receiver system. The IRRs of the two 2SB mixers were calculated from the $^{12}$CO ($J$ = 1--0) spectra at 115.271 GHz in the USB and HCO$^{+}$ ($J$ = 1--0) spectra at 89.182 GHz in the LSB from the W51 giant molecular cloud, resulting in $>$ 20 dB for one polarization and $>$ 12 dB for the other polarization.

\bigskip
The authors would like to thank Takafumi Kojima and Yasuhiro Abe for their contributions to this project. We are also grateful to Chieko Miyazawa, Akira Mori, and the entire staff of the Nobeyama Radio Observatory for their useful discussions and support. This study was supported in part by a Grant-in-Aid for Scientific Research on Priority Areas (15071205) from MEXT.

\end{document}